\documentclass[11pt]{article}
\usepackage{doublespace}
\usepackage{cite}
\usepackage{psfig}
\usepackage{latexsym}
\usepackage{array}

\def\be{\begin{equation}}
\def\ee{\end{equation}}
\def\bea{\begin{eqnarray*}}
\def\eea{\end{eqnarray*}}
\def\bean{\begin{eqnarray}}
\def\eean{\end{eqnarray}}

\def\lb{\linebreak}

\def\sig{\sigma}

\begin{document}
\begin{spacing}{2}

\title{Absence of Correlation between the Solar Neutrino Flux and \\
the Sunspot Number} 

\author{Guenther Walther \\
Dept. of Statistics, Stanford University, Stanford, CA 94305}
\date{}
\maketitle

\begin{abstract}
There exists a considerable amount of research
claiming a puzzling anti-correlation between the 
neutrino detection rate at the Homestake
experiment and indicators of solar activity
such as the sunspot number, giving rise to explanations involving
the hypothesis of a neutrino magnetic moment.
It is argued here that the claimed significant
anti-correlation is due to a statistical fallacy.
A proper test based on certain optimality criteria fails
to detect a significant time variation of the neutrino flux
in concert with the sunspot number, providing evidence that
the observations are consistent with no correlation
between the two series.
\end{abstract}

\newpage

Solar neutrinos are the only known particles to reach
Earth directly from the solar core and thus allow to test
directly the theories of stellar evolution and nuclear energy
generation \cite{ba}. A perceived anti-correlation between the
neutrino detection rate at the Homestake
experiment \cite{da} and indicators of solar activity
such as the sunspot number has been the object of a 
considerable amount of research \cite{bfp,bp,bu,bs,da,de,do,kr,ma,mc,oa}, 
yielding claims of statistically highly significant results.
Such time variations of the solar neutrino flux are not possible
in minimal standard electroweak theory and have motivated proposals
for solutions of the solar neutrino problem based upon
the hypothesis of a large neutrino magnetic moment \cite{ci,ok,vo,vo2}.

However, the standard tests for correlation used in the research
cited above require assumptions that
are usually not met in a time-series context, where these tests may
readily produce erroneous, highly significant results.
Figure~1 illustrates one aspect of this fallacy, which is often
ignored by statistics text books and therefore easily goes
unrecognized in scientific work: The top scatterplot shows
the first 100 of 109 typical independent observations $(X_1,Y_1),
\ldots,(X_{109},Y_{109})$ from a standard bivariate normal
distribution. The bottom scatterplot shows the 100 running
means of length 10, $(\frac{1}{10}\sum_{i=k}^{k+9} X_i,
\frac{1}{10}\sum_{i=k}^{k+9} Y_i),
k=1,\ldots,100$. The correlation
is visibly larger in the bottom plot. Indeed, Pearson's correlation
coefficient $r$ is $0.12$ for the top plot, and $0.30$ for the
bottom plot. However, the probability of obtaining  values of $|r|$
of at least the observed size is {\sl larger} for the situation
of the bottom plot (27\%) than
for that of the top plot (24\%), as can be verified by simulations!
 This example illustrates the fact that common
tests for correlation between two series tend to give erroneous,
highly significant results when there
is dependence within each of the two series, e.g. when the series exhibit
periodic behavior or are smoothed,
a commonly employed procedure either implicitly in the data collection
process or afterwards.

The sunspot numbers clearly have a strong dependence structure due
to the 11 year period of the sunspot cycle.
Table~1 shows how easily one is lead to an erroneous claim
of a significant correlation between the sunspot numbers and
an independent random series, this time using Spearman's
rank correlation coefficient $r_s$, another popular
measure of correlation: $X$ is taken to be the series of the 100
monthly sunspot numbers starting January 1970. $Y$ is a random walk with
independent Gaussian increments in row~1, and a 2-point and 
4-point running mean of independent standard Gaussian random variables in
rows 2 and 3, respectively. $Y$ was simulated $10^5$ times for each case,
and the columns give the relative frequency of
rejection of the null hypothesis of independence at nominal significance
levels $5\%$, $1\%$ and $0.1\%$, using the null
distribution of $r_s$ given in \cite{pf}. For example,
at the $1\%$ significance level one is led to the conclusion 
that there is a correlation with the random walk about 77\%
of the time! A comparison of rows~2 and 3 shows that a larger
degree of smoothing applied to one series makes the correlation
seemingly more significant, a fact that will be of importance below.
The effect described above is relevant quite generally for many 
tests of association or correlation,
such as the $\chi^2$ statistic for contingency tables,
or Kendall's tau statistic. It applies directly to those published results
on a perceived correlation between the sunspot number and the neutrino
flux that employ a smoothing of the neutrino flux. 

Furthermore, the assumptions
of these tests can also be violated in other important ways.
For example, tests for correlation using $r_s$ or Kendall's tau require
that the distribution of the components of at least one the two series
is invariant under permutations, which implies equal means and variances
of the measurements in that series. Row~4 of table~1 provides an
important example that violates this requirement: 
The neutrino flux is taken to be constant equal to 1 for a random
time which is distributed exponentially with mean 10 months,
then the flux equals 3 for a random time with the same
distribution, then it is set back to 1, etc. The flux is measured independently
each month with a standard Gaussian measurement error. Incidentally,
a typical simulation of this model looks even similar to the 
real neutrino data. 
Simulations of the flux from this model are uncorrelated with
the sunspot number and have no connection to the solar cycle
whatsoever. (There is nothing special about the exponential distribution
chosen: virtually any random or deterministic time will produce similar
results to the ones quoted in the following).
Still, row~4 shows that $r_s$ erroneously reports a correlation
at the 1\% level for 22.4\% of the simulations. 
Clearly, this test misinterprets a change
in the neutrino flux that is unrelated to the solar cycle as a
correlation with the solar cycle. One can reproduce this effect
with the all the tests employed in \cite{bfp,bp,bu,bs,da,de,do,kr,ma,mc,oa}.
This example makes clear that in this time series context, it is not
correct to interpret significant results of these tests as significant
evidence for a correlation with the solar cycle, even if no smoothing
of the neutrino data is employed.

One may ask whether these tests are at least providing evidence
for a time variation, not necessarily in concert with the solar cycle.
However, due to the unequal uncertainties in the neutrino measurements,
these tests are also not valid for testing whether the flux is constant:
For an illustration, let $x=(3,2,1,5)$ be a vector of four observations,
and $Y_1,\ldots,Y_4$ be four independent Gaussian random variables
with mean 0 and standard deviations $1,1,1$ and $4$. The $Y$'s represent
observations of a constant quantity with measurement error. In 7.0\% of $10^5$
simulations of the $Y$'s, the correlation $r_s$ between $x$ and $Y$
was equal to 1, whereas the table for the exact null distribution of $r_s$
gives a value of 4.17\% (see e.g. Table VIII in \cite{bd}). Similar results
obtain when the significance of the $\chi^2$- and $F$-statistics is evaluated
by randomly shuffling the data  \cite{bfp,bs}, as the distributions of
these statistics are not invariant under those permutations:
The best correlation (smallest value of $\chi^2$, resp. largest value of $F$)
is obtained by exactly one of the $4!=24$ permutations of the data,
yielding a significance level of $1/24=4.17\%$. However, this best
correlation was obtained in 10.3\% of the simulations.
While this effect seems to become less severe with more data or more equal
uncertainties, the example shows that these tests lack proper justification
and can produce invalid results. 
More importantly, when a modified test is used that accounts 
for the uncertainties in a proper way, then the highly significant results
reported for the neutrino data disappear:

The neutrino data that shall first be examined
are the 108 estimates $N_i$ of the neutrino
flux provided by the Homestake experiment \cite{da} up to run no.133, so that
$N_i = \mbox{flux($t_i$)} + \sig_i e_i,\ \ i=1,\ldots,108$.
Here flux($t$) denotes the neutrino flux at time $t$, which is possibly
time-varying.
The uncertainties $\sig_i$ given by the Homestake experiment
have recently been reanalyzed by the Homestake team, resulting in improved
uncertainties that have generously been made available
by Dr. Kenneth Lande~(private communication). 
The standardized measurement errors $e_i$
for the various runs are independent by the design of the experiment.
A test for correlation can now be developed
by examining how linear functions $a + b \cdot s_i$ of the monthly
sunspots numbers $s_i$ explain flux($t_i$), i.e. using regression techniques.
Under the null hypothesis of a constant neutrino flux, flux($t)=a$,
the distribution of the
scaled differences $d_i=(N_i-a)/\sig_i,\ i=1 \ldots 108$, is invariant under
permutations, which justifies the validity of a permutation test for the
statistic $T = \sum_{i=1}^{108} s_i d_i$. This statistic is
sensitive to trends in flux($t$) that vary in concert with the
$s_i$, and possesses certain optimality properties for this type
of problem \cite{mar}. $a$ was estimated by the standard estimate
$(\sum_{i=1}^{108} N_i/\sig_i^2)/\sum_{i=1}^{108} \sig_i^{-2}$.
The (improved) uncertainties provided by the Homestake experiment
were used in the same way as in \cite{bfp}, i.e. the test was done
using both `average errors' and `upper errors' for the $\sig_i$.
Using $10^4$ random permutations,
the test resulted in a two-tailed significance probability of 16.3\%
for average errors, and 10.4\% for upper errors.

As pointed out by a referee, it is informative to evaluate $T$ for
earlier stretches of the data, where highly significant correlations have
been reported: One obtains only marginally significant
results (significance around 2\%) for the data up to run no.108.
The same significance obtains for the stretch from run no.49 to run no.104,
after the result is adjusted by a factor of 10 due to favorable
`fishing' for a significant stretch as suggested in \cite{bp}.

A summary of these results also allows to put together to a coherent
picture the sometimes conflicting evidence reported in 
\cite{bfp,bp,bu,bs,da,de,do,kr,ma,mc,oa}:
The data up to run no.133 are clearly consistent with
a constant neutrino flux when tested against the 
alternative of a time variation in concert with the solar cycle,
according to a test with certain optimality properties for this problem.
The previously reported highly significant results in earlier stretches
of the data cannot be reproduced when the uncertainties
and the permutation argument are employed correctly. Only marginal evidence
for a time variation is found in these stretches.
In any case, it would not be correct to interpret
these results as evidence for a correlation
with the solar cycle. This allows to reconcile these findings with the
periodogram analysis in \cite{bp}, which shows no significant 11 yr component
in the data.
The reported improved correlation with smoother functions of the sunspot
numbers \cite{bp,de,ma,oa} is not surprising in light of the artifact exhibited
in the third paragraph.

I wish to thank Raymond Davis and Kenneth Lande for kindly
making the Homestake data available, Peter Sturrock
for valuable discussions, and a referee for criticism that
helped improve the paper. This work was supported by
the Air Force Office of Scientific Research and by NASA.

\begin{figure}[p]
\label{fig1}
\centerline{\psfig{figure=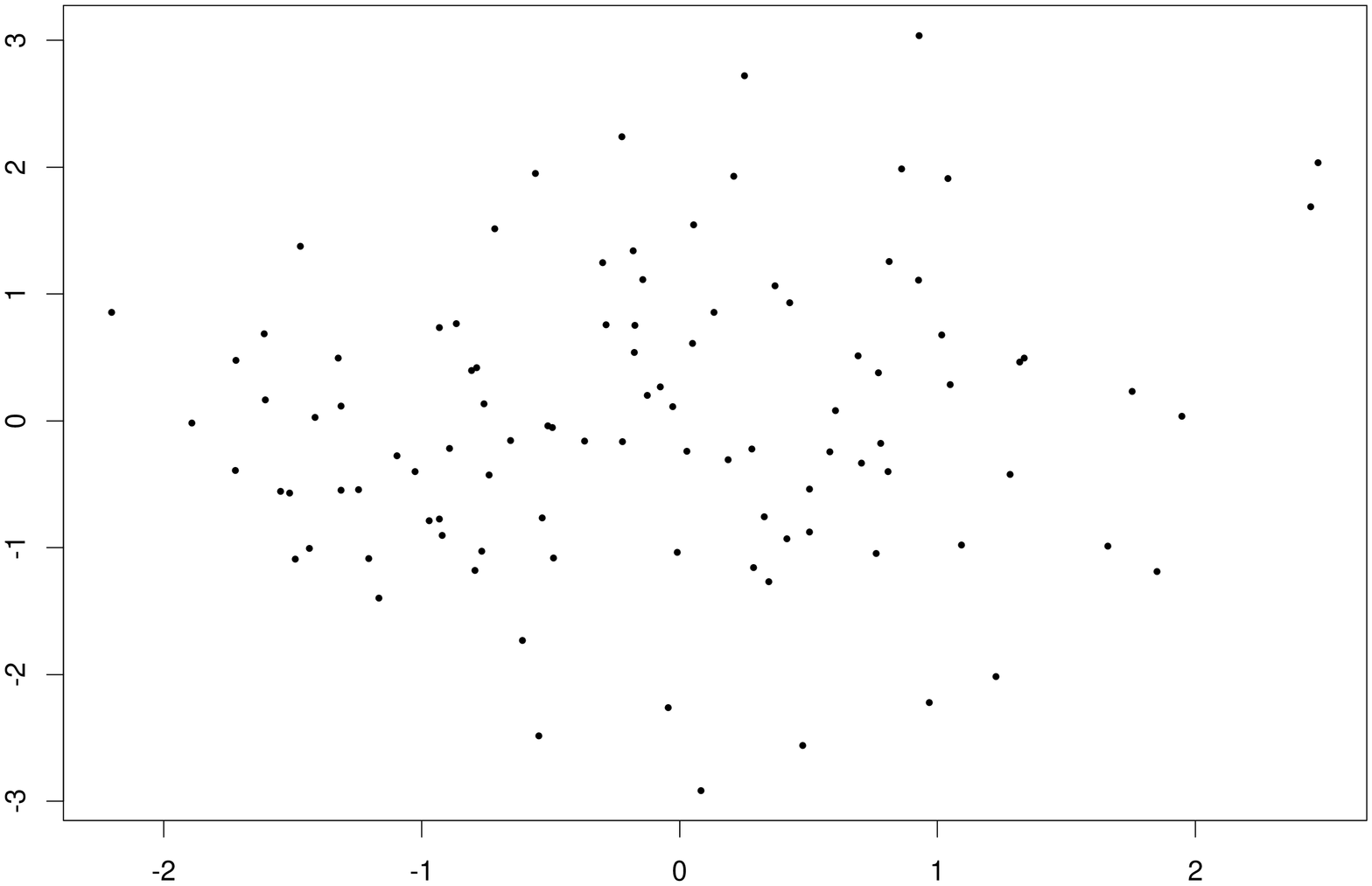,bbllx=20pt,bblly=11pt,bburx=592pt,bbury=781pt,width=5in,height=3.7in,angle=270}}
\centerline{\psfig{figure=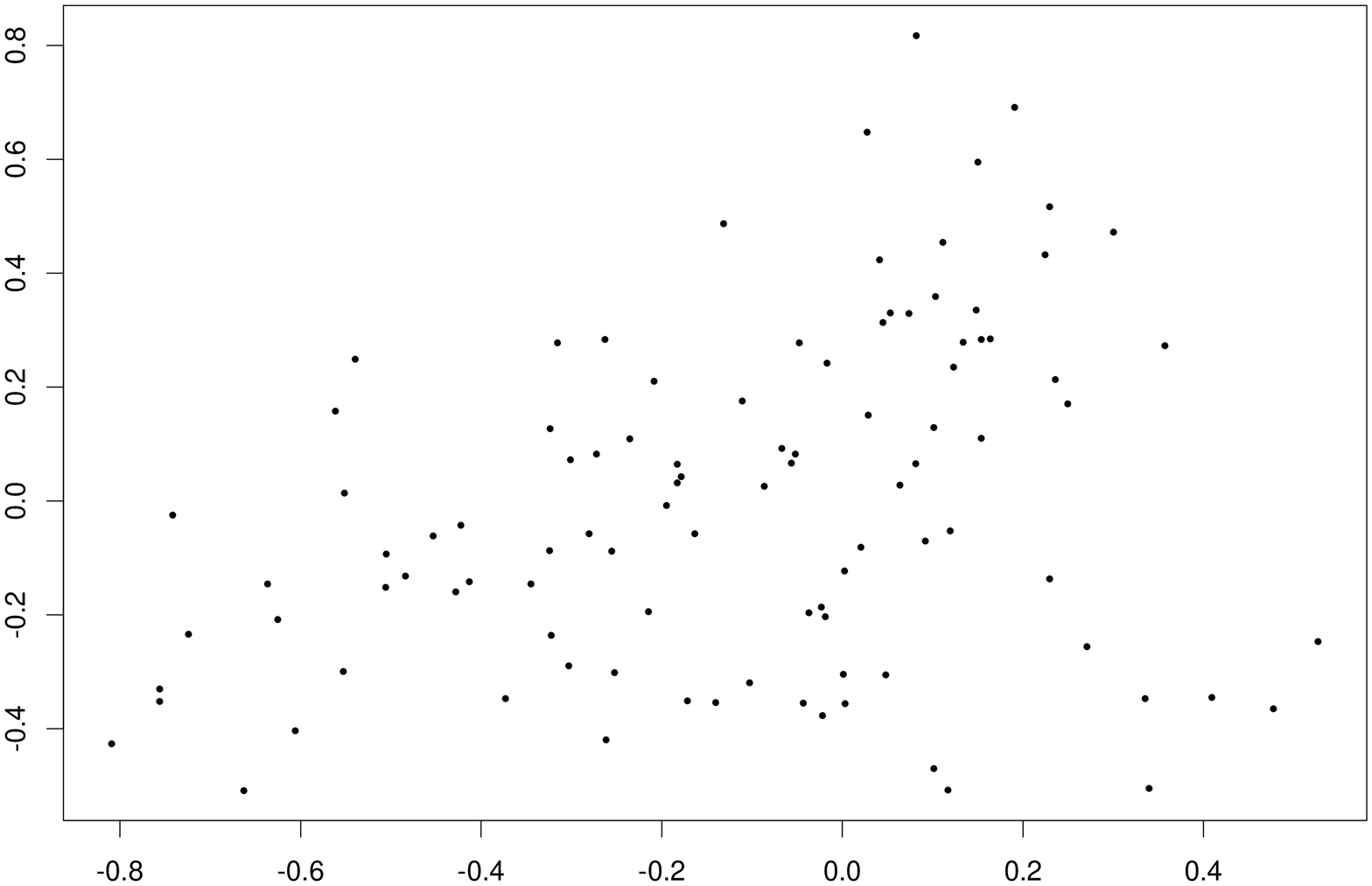,bbllx=20pt,bblly=11pt,bburx=592pt,bbury=781pt,width=5in,height=3.7in,angle=270}}
\caption{Top: Typical scatterplot of 100 independent standard bivariate
normal observations. Bottom: Running means of length 10. Pearson's correlation
coefficient is 0.12 for the top plot and 0.30 for the bottom plot.}
\end{figure}

\begin{table}[p]
\begin{tabular}{|c|m{2.5in}|r|r|r|} 
\hline
 &  & \multicolumn{3}{c|}{Relative frequency of rejection} \\
 &  & \multicolumn{3}{c|}{at nominal significance level} \\
 & \hspace{1cm} Time series & \hspace{0.5cm} 5\% \hspace{0.5cm} & 
   \hspace{0.5cm} 1\% \hspace{0.5cm} & \hspace{0.5cm} 0.1\% \hspace{0.5cm} \\
   \hline
1 & $X=$ sunspot numbers  \hspace{2.3cm} \lb 
  $Y_k=\sum_{i=1}^k Z_i$, $\ k=1,\ldots, 100$
  & 82.7\% \hspace{0.2cm} & 77.3\% \hspace{0.2cm} &  70.7\% \hspace{0.5cm} \\
  \hline
2 & $X=$ sunspot numbers  \hspace{2.3cm} \lb 
  $Y_k=\sum_{i=k}^{k+1} Z_i$, $\ k=1,\ldots, 100$
  & 15.3\% \hspace{0.2cm} & 6.1\% \hspace{0.2cm} &  1.7\% \hspace{0.5cm} \\
  \hline
3 & $X=$ sunspot numbers  \hspace{2.3cm} \lb 
  $Y_k=\sum_{i=k}^{k+3} Z_i$, $\ k=1,\ldots, 100$
  & 30.0\% \hspace{0.2cm} & 17.5\% \hspace{0.2cm} &  8.4\% \hspace{0.5cm} \\
  \hline
4 & $X=$ sunspot numbers  \hspace{2.3cm} \lb
  $Y_k=1+Z_k\ $ if $T_{2i} < k \leq T_{2i+1}$, \hspace{1cm} \lb
  $Y_k=3+Z_k$ else; $\ k=1,\ldots, 100$
  & 35.7\% \hspace{0.2cm} & 22.4\% \hspace{0.2cm} &  11.7\% \hspace{0.5cm} \\
  \hline
\end{tabular}
\caption{Relative frequencies of rejection of the
null hypothesis of independence at various nominal significance levels 
in a Monte Carlo study using the nominal null distribution of Spearman's
correlation coefficient. $X$ is the series of the 100 monthly sunspot numbers 
starting in January 1970. The $Z_i$ are independent standard normal
random variables. $T_i$ is the sum of the first $i$ terms of a sequence of
independent exponential random variables with mean 10 months.}
\end{table}

\end{spacing}

\bigskip

\begin{thebibliography}{99}

\bibitem{ba} J.N. Bahcall, {\bf Neutrino Astrophysics} (Cambridge Univ. Press,
Cambridge, 1989).
\bibitem{bfp} J.N. Bahcall, G.B. Field, and W.H. Press, 
Astrophys. J. Lett. {\bf 320} L69 (1987).
\bibitem{bp} J.N. Bahcall and W.H. Press, Astrophys. J. {\bf 370} 730 (1991).
\bibitem{bs} J.W. Bieber, D. Seckel, T. Stanev, and G. Steigman,
Nature {\bf 348} 407 (1990).
\bibitem{bu} D. Basu, Solar Physics {\bf 81} 363 (1982).
\bibitem{da} R. Davis Jr., Nucl. Phys. B (Proc. Suppl.) {\bf 48} 284 (1996).
\bibitem{de} P.H. Delache, V. Gavryusev, E. Gavryuseva, F. Laclare,
C. Regulo, and T. Roca Cortes, Astrophys. J. {\bf 407} 801 (1993).
\bibitem{do} L.I. Dorman and A.W. Wolfendale, J. Phys. G. Nuclear Part.
Phys. {\bf 17} 769 (1991).
\bibitem{kr} L.M. Krauss, Nature {\bf 348} 403 (1990).
\bibitem{ma} S. Massetti and M. Storini, Solar Physics {\bf 148} 173 (1993).
\bibitem{mc} R.L. McNutt Jr., Science {\bf 270} 1635 (1995).
\bibitem{oa} D.S. Oakley, H.B. Snodgrass, R.K. Ulrich, and T.L. VanDeKop, 
Astrophys. J. Lett. {\bf 437} L63 (1994).
\bibitem{ci} A. Cisneros, Astrophys. Space Sci. {\bf 10} 87 (1971).
\bibitem{ok} L.B. Okun, Soviet J. Nucl. Phys. {\bf 44} 546 (1986).
\bibitem{vo} M.B. Voloshin, M.I. Vysotskii, and L.B. Okun, 
Soviet J. Nucl. Phys. {\bf 44} 440 (1986).
\bibitem{vo2} M.B. Voloshin and M.I. Vysotskii, 
Soviet J. Nucl. Phys. {\bf 44} 544 (1986).
\bibitem{pf} W.H. Press, B.P. Flannery, S.A. Teukolsky, and W.T. Vetterling, 
{\bf Numerical Recipes in C: The Art of Scientific Computing}
(2nd ed., Cambridge Univ. Press, New York, 1992).
\bibitem{bd} P.J. Bickel, K.A. Doksum, {\bf Mathematical Statistics,
Basic Ideas and Selected Topics} (Holden-Day, Oakland, 1977).
\bibitem{mar} J.S. Maritz, {\bf Distribution-Free Statistical Methods}
(Chapman \& Hall, London, 1995).
\end{thebibliography}
\end{document}